\documentclass{emulateapj}

\usepackage{times}
\usepackage{graphicx,color}
\usepackage{hyperref}
\usepackage{ulem}

\def\Fermi{\textit{Fermi}}
\def\gam{$\gamma$}
\def\deg{$^{\circ}$}
\def\VelaX{Vela$-$X}
\definecolor{azur}{rgb}{0.05,0.05,0.95}

\shorttitle{The \VelaX\ PWN revisited with 4 years of $Fermi$-LAT observations}
\shortauthors{Grondin et al., 2013}

\begin{document}

\title{The Vela$-$X Pulsar Wind Nebula revisited with 4 years of $Fermi$ Large Area Telescope observations}

\author{
M.-H.~Grondin\altaffilmark{1,2,3}, 
R.~W.~Romani\altaffilmark{4},
M.~Lemoine-Goumard\altaffilmark{5,6}, 
L.~Guillemot\altaffilmark{7}
A.~K.~Harding\altaffilmark{8}, 
T.~Reposeur\altaffilmark{5}, 
}
\altaffiltext{1}{Max-Planck-Institut f\"ur Kernphysik, P.O. Box 103980, D 69029
Heidelberg, Germany}
\altaffiltext{2}{Now at CNRS, IRAP, F-31028 Toulouse cedex 4, France / GAHEC, Universit\'e de Toulouse, UPS-OMP, IRAP, Toulouse, France}
\altaffiltext{3}{email: mgrondin@irap.omp.eu}
\altaffiltext{4}{W. W. Hansen Experimental Physics Laboratory, Kavli Institute for Particle Astrophysics and Cosmology, Department of Physics and SLAC National Accelerator Laboratory, Stanford University, Stanford, CA 94305, USA}
\altaffiltext{5}{Universit\'e Bordeaux 1, CNRS/IN2p3, Centre d'\'Etudes Nucl\'eaires de Bordeaux Gradignan, 33175 Gradignan, France}
\altaffiltext{6}{Funded by contract ERC-StG-259391 from the European Community}
\altaffiltext{7}{Max-Planck-Institut f\"ur Radioastronomie, Auf dem H\"ugel 69, 53121 Bonn, Germany}
\altaffiltext{8}{NASA Goddard Space Flight Center, Greenbelt, MD 20771, USA}
\begin{abstract}
The Vela supernova remnant is the closest supernova remnant to Earth containing an active pulsar, the Vela pulsar (PSR B0833$-$45). This pulsar is the archetype of the middle-aged pulsar class and powers a bright pulsar wind nebula (PWN), \VelaX, spanning a region of 2\deg\ $\times$ 3\deg\ south of the pulsar and observed in the radio, X-ray and very high energy \gam-ray domains. The detection of the \VelaX\ PWN by the \Fermi\ Large Area Telescope (LAT) was reported in the first year of the mission. Subsequently, we have re-investigated this complex region and performed a detailed morphological and spectral analysis of this source using 4 years of \Fermi-LAT observations. This study lowers the threshold for morphological analysis of the nebula from 0.8 GeV to 0.3 GeV, allowing inspection of distinct energy bands by the LAT for the first time. We describe the recent results obtained on this PWN and discuss the origin of the newly detected spatial features.
\end{abstract}

\keywords{Gamma rays: general - ISM: individual objects: Vela-X - pulsars: general Ð pulsars: individual (Vela, PSR~J0835$-$4510)}
\section{Introduction}
The supernova remnant (SNR) G263.9-3.3 (aka the Vela SNR) is the closest composite SNR to Earth containing an active pulsar, the Vela pulsar (PSR~B0833$-$45) and is therefore studied in great detail across the electromagnetic spectrum. Located at a distance of only $D$~=~290~pc \citep{Caraveo2001, Dodson2003}, the Vela pulsar has a characteristic age of $\tau_c$~=~11~kyr, a spin period of $P$~=~89~ms and a spin-down power of $\dot{E}$~=~7~$\times$~10$^{36}$ erg~s$^{-1}$. First discovered as a radio loud pulsar \citep{Large1968}, its  pulsations were successively detected in high energy (HE) \gam-rays \citep{Thompson1975}, optical \citep{Wallace1977} and X-rays \citep{Ogelman1993}. Recent \gam-ray observations by the \Fermi\ Large Area Telescope (LAT) have confirmed its detection above 20~MeV and enabled a more detailed study of its \gam-ray properties than possible with the previous missions SAS-II, COS-B and CGRO-EGRET \citep[][ and references therein]{Kanbach1994}. These observations reveal a magnetospheric emission over 80\% of the pulsar period, and a strong and complex phase dependence of the \gam-ray spectrum, in particular in the peaks of the light curve \citep{Vela1, Vela2}.

The 8\deg-diameter Vela SNR is also known to host several regions of non-thermal and diffuse radio emission labelled Vela$-$X, Vela$-$Y and Vela$-$Z \citep{Rishbeth1958}. The brightest one ($\sim$~1000~Jy), \VelaX\ , spans a region of 2\deg~$\times$~3\deg\ (referred to as the ``$halo$'') surrounding the Vela pulsar and shows a filamentary structure. In particular, the brightest radio filament has an extent of 45$^\prime$~$\times$~12$^\prime$ and is located south of the pulsar. The flat spectrum of \VelaX\ with respect to Vela$-$Y and Vela$-$Z and its high degree of radio polarization have led to strong presumptions that \VelaX\ is the pulsar wind nebula (PWN) associated with the energetic and middle-aged Vela pulsar \citep{Weiler1980}. The rotational energy of the pulsar is dissipated through a magnetized wind of relativistic particles. A PWN forms at the termination shock resulting from the interaction between the relativistic wind and the surrounding material, e.g. the supernova ejecta \citep{Gaensler2006}.

Following its radio discovery, the \VelaX\  region has been intensively observed at every wavelength. X-ray observations by $ROSAT$ have unveiled a diffuse and nebular emission (with an extent of 1.5$^\prime$~$\times$~0.5$^\prime$) coincident with the bright radio filament and referred to as the ``$cocoon$'' \citep{Markwardt1995}. The \VelaX\ region has been significantly detected up to 0.4~MeV by $OSSE$ with a spectrum consistent with the  $E^{-1.7}$ spectrum seen between optical and X-rays \citep{DeJager1996}. High resolution $Chandra$ observations have enabled the detection of bright and compact non-thermal  X-ray emission composed of two toroidal arcs (17$^\prime$$^\prime$ and 30$^\prime$$^\prime$ away from the pulsar) and a 4$^\prime$-long collimated ``jet''-like structure \citep{Helfand2001}. Finally, very high energy (VHE) \gam-ray observations by the H.E.S.S. telescopes have revealed bright emission spatially coincident with the cocoon, whose spectrum peaks at $\sim$~10 TeV \citep{Aharonian2006}. This detection has confirmed the non-thermal nature of the cocoon; however, the relativistic particle population responsible for the X-ray and TeV emission can hardly account for the halo structure observed in radio.

At this time, several scenarios have been proposed to reconcile multi-wavelength data. \cite{Horns2006} proposed a hadronic model, in which the VHE \gam-ray emission is explained by proton-proton interactions inside the cocoon followed by neutral pion decays. In parallel, \cite{DeJager2008} suggested the existence of two electron populations in \VelaX : a young population that produces the narrow cocoon seen in X-rays and at VHE, and a relic one responsible for the extended halo observed in radio. According to this model, significant emission from the halo should be detectable in the \Fermi-LAT energy range. An alternative scenario was recently proposed by \cite{Hinton2011}, which explains the observations by diffusive escape of particles in the extended halo structure.
 
The first HE \gam-ray detection of the \VelaX\ PWN by the \Fermi-LAT was reported in the first year of the mission. The source is significantly extended (with an extension of $\sigma_{Disk}$ = 0.88\deg~$\pm$~0.12\deg\ assuming a uniform disk hypothesis) and its spectrum is well reproduced with a simple power law having a soft index ($\Gamma$~$\sim$~2.41~$\pm$~0.09$_{stat}$~$\pm$~0.15$_{syst}$) in the 0.2~--~20~GeV energy range \citep{VelaX1}. The detection of \VelaX\ in the 0.1~--~3~GeV energy range was also reported by the AGILE Collaboration \citep{Pellizzoni2010}. 

\cite{Abramowski2012} recently reported the detection of faint TeV emission spatially coincident with the radio halo, in addition to the bright emission already reported and matching the X-ray emission. This new result challenges the simple interpretation of a young electron population being responsible for the X-ray and VHE emission.

We have re-investigated this complex region  in HE \gam-rays and performed a detailed morphological and spectral analysis of this source using 4 years of \Fermi-LAT observations. The energy range for morphological analysis is extended down to 0.3 GeV allowing the  first study of energy-dependent morphology by the LAT. In this paper, we report the results of this analysis and discuss the main implications of the new spectrally-resolved spatial information in the context of the theoretical models described above. In particular, we discuss the possible interpretation of the energy-dependent morphology brought to light with 4 years of \Fermi-LAT observations.

\section{LAT description and observations} \label{section:observations}
The LAT is a \gam-ray telescope that detects photons by conversion into electron-positron pairs and operates in the energy range from 20 MeV to greater than 300 GeV. Details of the instrument and data processing are given in \cite{Atwood 2009}. The on-orbit calibration is described in \cite{Abdo2009b} and \cite{Ackermann2012}.

The following analysis was performed using 48 months of data collected starting 2008 August 4, and extending until 2012 August 4 within a 15\deg\ $\times$ 15\deg\ region around the position of the Vela pulsar. 

Only \gam-rays in the from the Pass 7 ``Source'' class were selected from this sample. This class corresponds to a good compromise between the number of selected photons and the background rate. We have used the P7SOURCE$\_$V6 Instrument Response Functions (IRFs) to perform the following analyses. We excluded photons with zenith angles greater than 100$^{\circ}$ to reduce contamination from secondary \gam-rays originating in the Earth's atmosphere \citep{Abdo2009b}. 

\section{Timing analysis of the pulsar PSR J0835-4510} \label{section:timing}

The Vela pulsar is the brightest steady point source in the \gam-ray sky, with pulsed photons observed up to 25 GeV, and is located within the \VelaX\ PWN. Previous analyses of its \gam-ray properties using \Fermi-LAT observations have shown that magnetospheric emission is observed over 80\% of the pulsar period \citep{Vela2}. 

The detailed study of \VelaX\ requires working in the off-pulse window of the Vela pulsar light curve (i.e. 20\% of the pulsar period) to avoid contamination from the pulsed emission. Because the Vela pulsar exhibits substantial timing irregularities, phase assignment generally requires a contemporary ephemeris. To perform the following analysis, \gam-ray photons were phase-folded using an accurate timing solution derived from \Fermi-LAT observations. The Vela pulsar is extremely bright in \gam-rays, and its continuous observations by the LAT since the beginning of the mission enables us to directly construct regular times of arrival (TOAs) that are then used to generate a precise pulsar ephemeris \citep{Ray2011}. 

The Vela pulsar experienced a glitch, i.e. a large jump in rotational frequency,  near MJD 55428. To avoid any contamination from the Vela pulsar in the analysis of its PWN, data between MJD 55407 and MJD 55429 were excluded from the dataset, and two timing solutions (pre- and post-glitch) were used to phase-fold the \gam-ray photons. The pre-glitch ephemeris was built using 198 TOAs  covering the period from the beginning of the science phase of the \Fermi\ mission (2008 August 04) to the glitch, while the post-glitch timing solution was built using 197 TOAs from the glitch to 2012 August 04. For both timing solutions, we fit the \gam-ray TOAs to the pulsar rotation frequency and first five derivatives. The fit further includes 10 harmonically related sinusoids, using the ÒFITWAVESÓ option in the TEMPO2 package \citep{Hobbs2006}, to flatten the timing noise. The post-fit rms is 91.3 $\mu$s and 97.7 $\mu$s (i.e. ~0.1\% of the pulsar phase) for the pre- and post-glitch ephemeris respectively. These timing solutions will be made available through the \Fermi\ Science Support Center (FSSC)\footnote{Pulsar timing models: http://fermi.gsfc.nasa.gov/ssc/data/access/lat/ephems/}. We define phase 0 for the model based on the fiducial point from the radio timing observations, which is the peak of the radio pulse at 1.4 GHz.

\begin{figure}
\begin{center}
\includegraphics[width=0.48\textwidth]{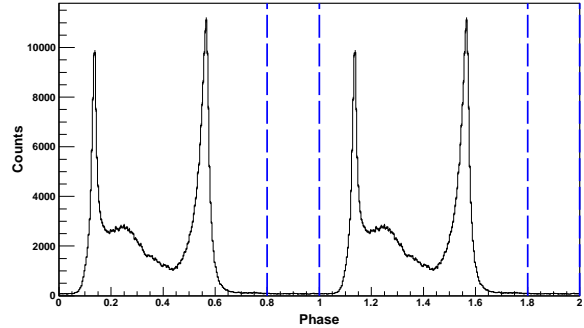}
\caption{Gamma-ray light curve of the Vela pulsar in the 0.1~--~300~GeV energy range using events in a 1\deg\ radius around the Vela pulsar position. The binning of the light curve is 0.004 in pulsar phase. The main peak of the radio pulse seen at 1.4 GHz is at phase 0. Two cycles are shown. The off-pulse window (shown by dashed blue lines) used for the analysis of the \VelaX\ PWN was defined between 0.8 and 1.0 of the pulsar phase.}
\label{fig:Phaso}
\end{center}
\end{figure}

Pulse phases were assigned to the LAT data using the \Fermi\ plug-in provided by the LAT team and distributed with TEMPO2. Only \gam-ray photons in the  0.8~--~1.0 pulse phase interval, corresponding to the off-pulse window, were selected and used for the spectral and morphological analysis presented in the following sections. Figure~\ref{fig:Phaso} shows the \gam-ray light curve of the Vela pulsar obtained in the 0.1~--~300~GeV energy range using events in a 1\deg\ radius around the position of the Vela pulsar and the definition of the off-pulse window (blue dashed lines). It is worth noting that this phase interval was chosen to be narrower than the one used in the previous \Fermi-LAT analysis to avoid any contamination from the Vela pulsar, which was estimated to be $\sim$~6\% in the 0.7~--~0.8 phase interval~\citep{VelaX1}. 

\section{Analysis of the Vela$-$X PWN}\label{section:analysis}
The spatial and spectral analysis of the \gam-ray data was performed using two different tools, $\mathtt{gtlike}$ and $\mathtt{pointlike}$. $\mathtt{gtlike}$ is a maximum-likelihood method \citep{Mattox1996} implemented in the Science Tools distributed by the FSSC. $\mathtt{pointlike}$ is an alternate binned likelihood technique, optimized for characterizing the extension of a source (unlike $\mathtt{gtlike}$), that has been extensively tested against $\mathtt{gtlike}$ \citep{Kerr2011, Lande2012}. These tools fit a model of the region, including sources, residual cosmic-ray, extragalactic and Galactic backgrounds, to the data.

In the following analysis, the Galactic diffuse emission is modeled using the standard model \textit{gal\_2yearp7v6\_v0.fits}. The residual charged particles and extragalactic radiation are described by a single isotropic component with a spectral shape described by the file \textit{iso\_p7v6source.txt}. The models and their detailed description are released by the LAT Collaboration\footnote{These models are available at : http://fermi.gsfc.nasa.gov/ssc/data/access/lat/BackgroundModels.html}.  

Sources within 10$^{\circ}$ of the Vela pulsar are extracted from the Second \Fermi-LAT Catalog \citep{SecondCat} and used in the likelihood fit. 
The nearby bright and extended SNRs Puppis~A and Vela~Jr are modeled with their best-fit models, i.e. a uniform disk of radius 0.38\deg\ for Puppis~A \citep{PuppisA} and the template of the TeV emission as seen with H.E.S.S. for Vela~Jr \citep{VelaJr}. The spectral parameters of sources closer than 3$^{\circ}$ to \VelaX\ are left free, while the parameters of all other sources are fixed at the values from \cite{SecondCat}.  Due to the longer integration time of our analysis (48 months vs. 24 months in the catalog) and the overwhelming brightness of the Vela pulsar in the full phase interval, the appearance of additional sources in our region of interest is expected. These sources, denoted with the identifiers BckgA and BckgB, were also considered in the analysis of SNR Puppis~A and were fit at the following positions : BckgA at $\alpha(\rm{J2000})=125.77$ \deg, $\delta(\rm{J2000})=$-$42.17$\deg\ with a 68\% error radius of 0.06\deg; BckgB at $\alpha(\rm{J2000})=128.14$\deg, $\delta(\rm{J2000})=$-$43.39$\deg\ with a 68\% error radius of 0.05\deg. More details on these sources are available in \cite{PuppisA}.

\subsection{Morphology}\label{subsection:morpho}

\begin{deluxetable*}{lccccccc}
\tablecaption{Centroid and extension fits to the LAT data for \VelaX\ using $\mathtt{pointlike}$ for $front$ events above 0.3 GeV. \label{table:centroid_pointlike} }
\tablewidth{0pt}
\tablehead{ 
\colhead{Spatial Model} & \colhead{R.A. ($^\circ$)} & \colhead{Dec. ($^\circ$) } & $\sigma_1$ ($^\circ$) & $\sigma_2$ ($^\circ$)& P.A. ($^\circ$) &  \colhead{$2 (logL_{1}- logL_{0})$*} & \colhead{Add. d.o.f.**} 
}
\startdata
Point Source 			& 128.40	& $-$45.54 	&  				& 				&				& 237 	& 4 \\
Gaussian				& 128.49 	& $-$45.38	& 0.80 $\pm$ 0.05	& 				& 				& 453 	& 5 \\
Disk					& 128.46	& $-$45.37	& 1.27 $\pm$ 0.10	& 				&				& 427 	& 5 \\
Elliptical Gauss  		& 128.40  & $-$45.40  	& 0.46 $\pm$ 0.05 	& 1.04 $\pm$ 0.08	& 40.3 $\pm$ 0.3    			& 481 	& 7 \\
\vspace{0.2cm} Elliptical Disk  			& 128.53  & $-$45.38   & 0.85 $\pm$ 0.05 	& 1.77 $\pm$ 0.09	& 39.6  $\pm$ 4.2  			& 476 	& 7 \\
HESS 				& 		& 		& 				& 				&				& 316  	& 2 \\
Radio 				&  		& 		& 				& 				&				& 431 	& 2 \\
Southern radio wing		&		&		&				&				&				& 401	& 2 \\
Split radio model 		& 		& 		& 				& 				&				& 447	& 4 \\
\enddata
\tablenotetext{*}{$L_{1}$ and $L_{0}$ are defined as the likelihood values corresponding to the fit of the spatial model described in the first column plus the background model and the fit of the background model only (null hypothesis). } 
\tablenotetext{**}{Add. d.o.f. : additional degrees of freedom.} 
\end{deluxetable*}

Previous analysis of the \VelaX\ PWN using 11 months of \Fermi-LAT data have shown that the source is significantly extended above 0.8 GeV, with an extension of $\sigma_{Disk}$ = 0.88\deg~$\pm$~0.12\deg\ assuming a uniform disk  \citep[hereafter labelled ``$Disk11m$''; ][]{VelaX1}.  

The increasing statistics and the improvement of the IRFs with respect to \cite{VelaX1} allow a more detailed study of the source and the use of a lower energy threshold of 0.3~GeV. 

To study the morphology of an extended source, a major requirement is to have the best possible angular resolution. Consequently, we restrict the LAT data set to $front$ events only, i.e. events which convert in the thin layers of the tracker, which benefit from higher angular resolution\footnote{For more information, please see the \Fermi-LAT performance page : http://www.slac.stanford.edu/exp/glast/groups/canda/lat$\_$Performance.htm} \citep{Atwood 2009}.

\begin{figure}
\begin{center}
\includegraphics[width=0.49\textwidth]{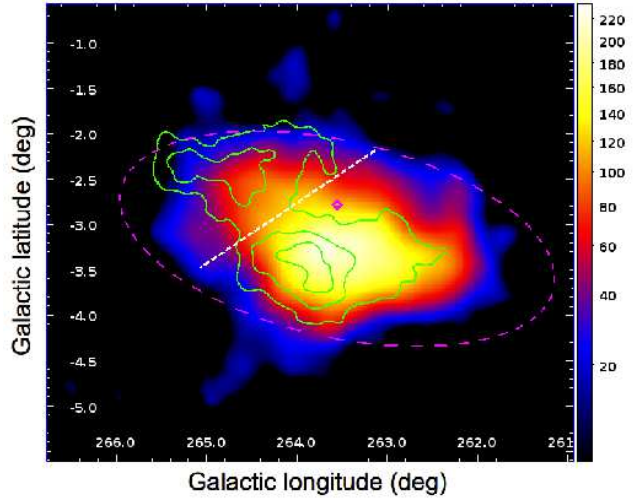}
\caption{Gamma-ray TS map of the \VelaX\ PWN in the 0.3 -- 100 GeV energy range (using $front$ events only, Galactic coordinates). The 61 GHz WMAP radio contours (0.80, 0.95 and 1.1 mK) are overlaid for comparison. The dashed white line shows the division of the radio template in two halves. The dashed magenta ellipse shows the best-fit morphological model, i.e. the elliptical Gaussian (99\% containment). The position of the Vela pulsar is marked with a magenta diamond. The color-coding is represented on a square-root scale.}
\label{fig:TSmap1}
\end{center}
\end{figure}

Figure~\ref{fig:TSmap1} presents the \Fermi-LAT Test Statistic (TS) map of \gam-ray emission around the \VelaX\ PWN above 0.3~GeV using $front$ events only. The TS is defined as twice the difference between the likelihood $L_1$ obtained by fitting a source model plus the background model to the data, and the likelihood $L_0$ obtained by fitting the background model only : TS = 2($log L_1$ $-$ $logL_0$).  This skymap contains the TS value for a point source at  each map location, thus giving a measure of the statistical significance for the detection  of a \gam-ray source in excess of the background. The diffuse Galactic and isotropic emission, as well as nearby sources are included in the background model and subtracted from the map. 

We used $\mathtt{pointlike}$ to measure the source extension using five different spatial hypotheses: a point source, a uniform disk hypothesis, a Gaussian distribution, an elliptical Gaussian distribution and an elliptical disk \citep{Lande2012} assuming a power-law spectrum. The results of the extension fits and the improvement of the TS when using spatially extended models are summarized in the first half of Table~\ref{table:centroid_pointlike}, along with the number of additional degrees of freedom with respect to the null hypothesis.

The improvement of the likelihood fit between  a Gaussian distribution and the point-source hypothesis\footnote{The formula used to derive the significance of an improvement when comparing two different spatial models with different numbers of degrees of freedom is extracted from Particle Data Group \citep{Beringer2012}.} (difference in TS of 216, which corresponds to an improvement at a $\sim$15$\sigma$ level) supports a significantly extended source. The best-fit model in the 0.3~--~100~GeV energy range is obtained for an elliptical Gaussian distribution. This best-fit model represents a 5$\sigma$ improvement with respect to a symmetric Gaussian distribution ($\Delta$TS$=28$ for two additional degrees of freedom), which means that the source is also significantly elongated. The best-fit center of gravity of the emission region is (R.A., Dec.) =  (128.40\deg\ $\pm$ 0.05\deg, $-$45.40\deg $\pm$ 0.05\deg). The best-fit width along the major axis is 1.04\deg\ $\pm$ 0.09\deg, while the best-fit intrinsic width along the minor axis is 0.46\deg\ $\pm$ 0.05\deg. The major axis of the fitted distribution is at a position angle (P.A.) of 40.3\deg\ $\pm$ 4.0\deg. 

\begin{figure*}
\begin{center}
\includegraphics[width=0.4\textwidth]{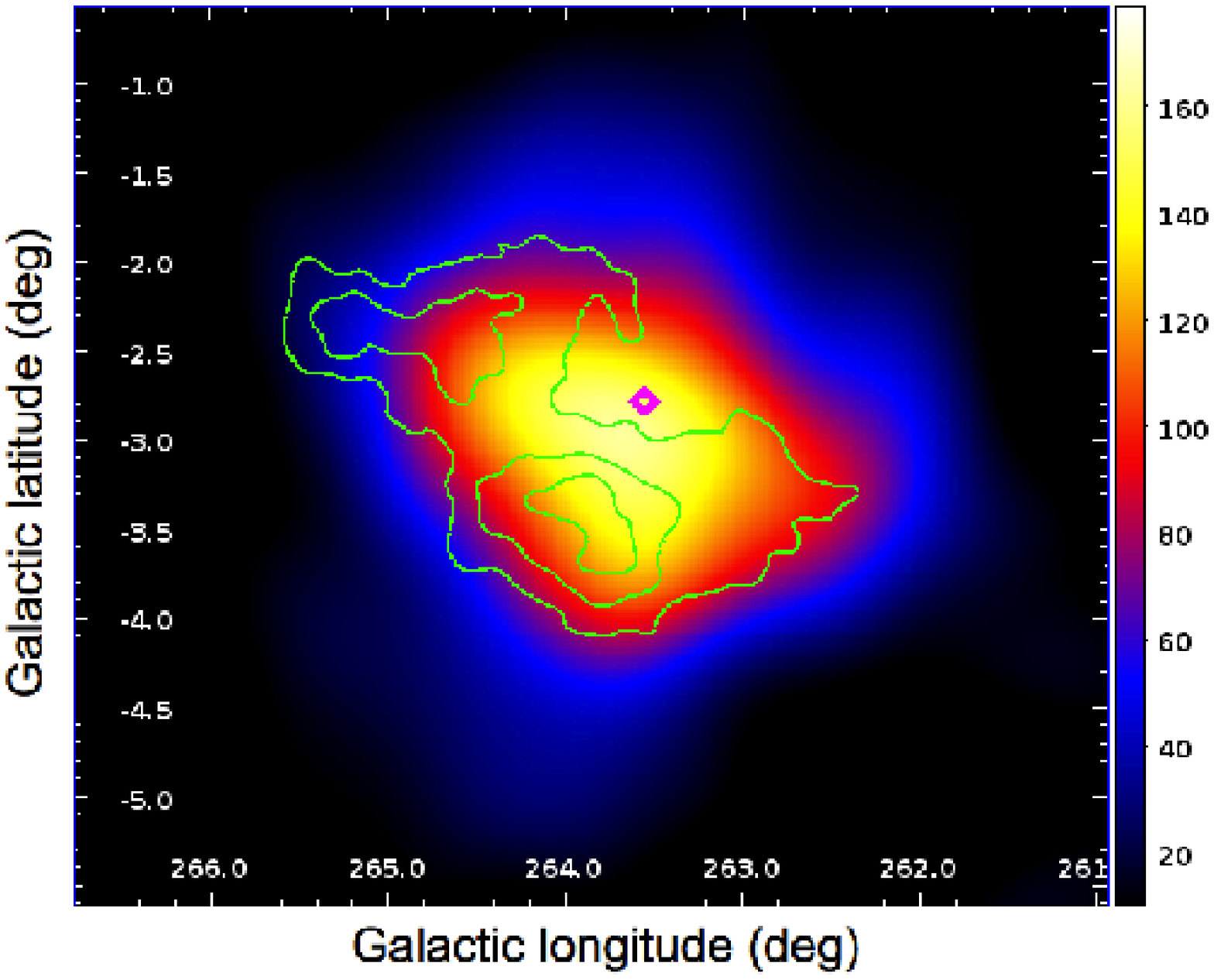}\includegraphics[width=0.4\textwidth]{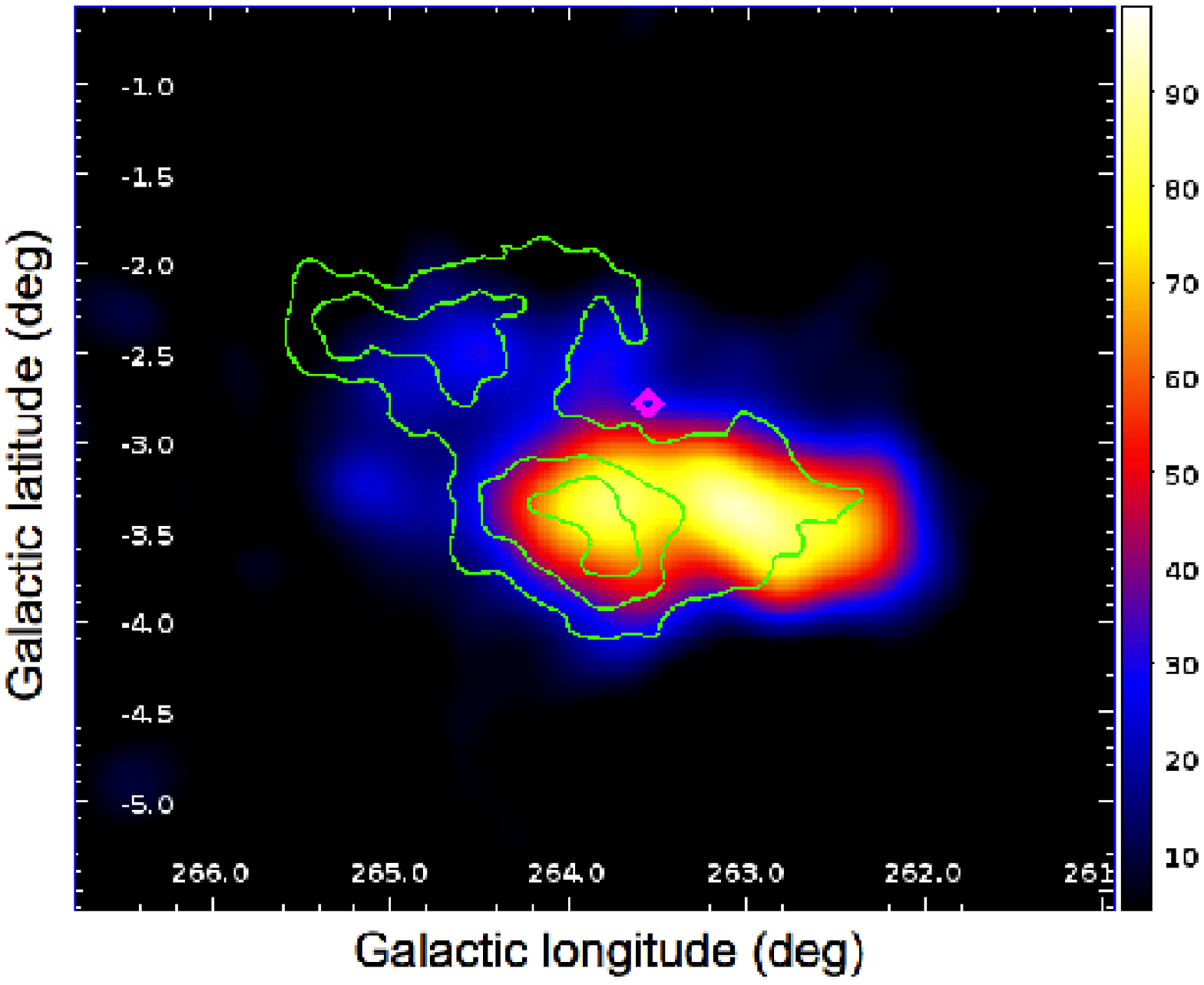}
\includegraphics[width=0.4\textwidth]{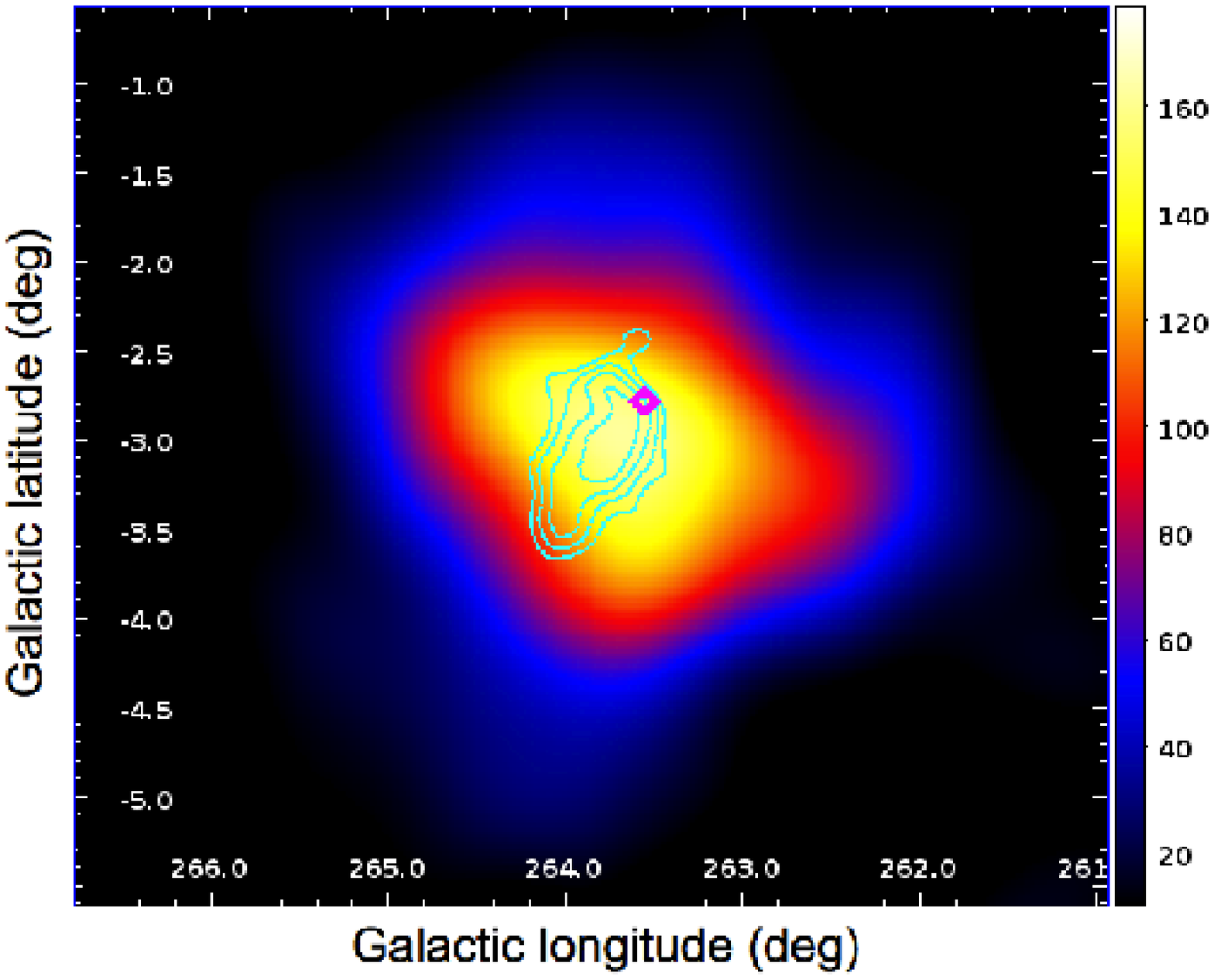}\includegraphics[width=0.4\textwidth]{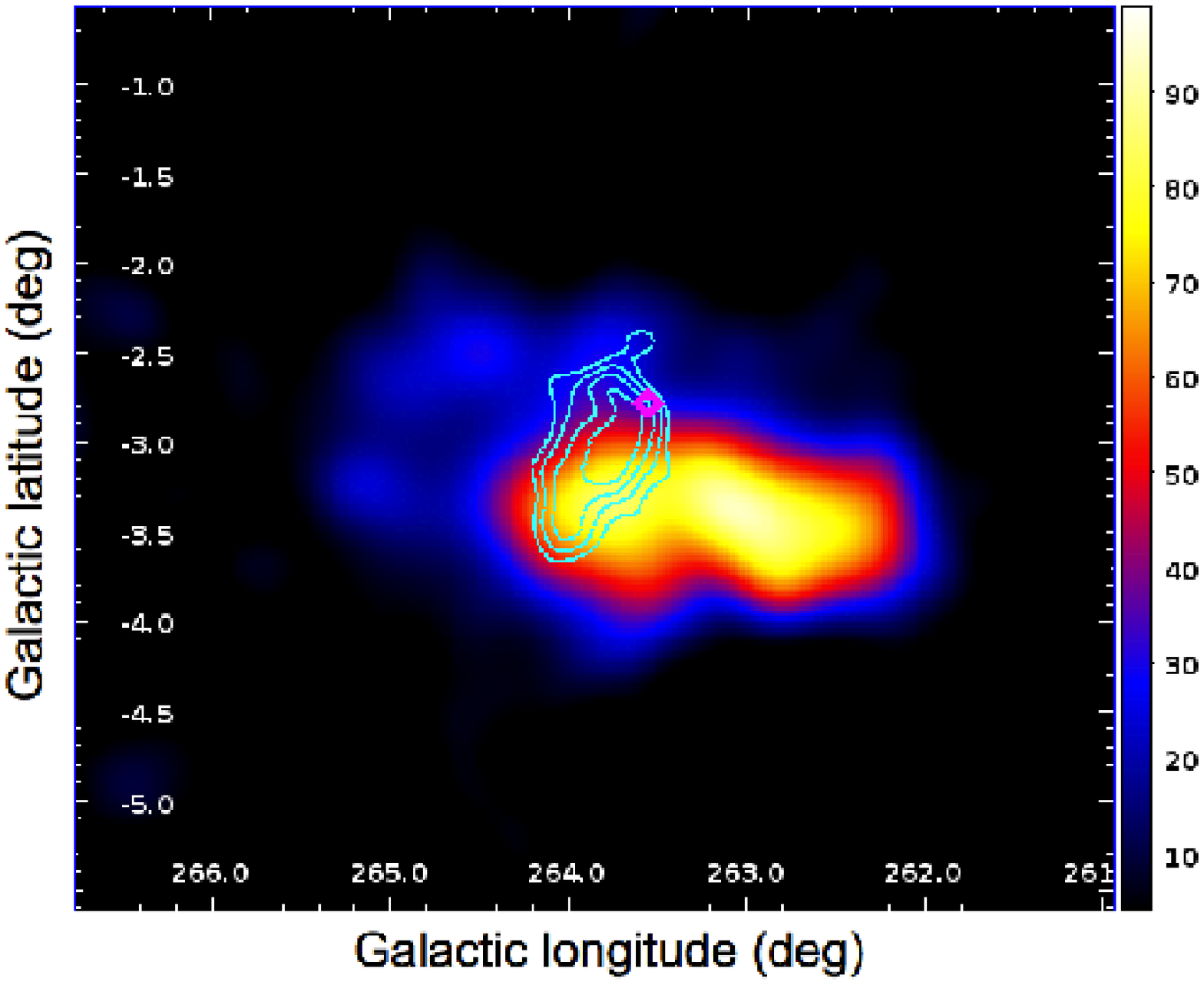}
\caption{\Fermi-LAT TS maps  of the \VelaX\ PWN in the 0.3 -- 1.0 (left) and 1.0 -- 100.0 GeV (right) energy ranges, (using $front$ events only, Galactic coordinates). The contours of the WMAP 61~GHz radio (in green, top row) and TeV emission \citep[in light blue, bottow row][]{Aharonian2006} are overlaid for comparison. The position of the Vela pulsar is marked with a magenta diamond. }
\label{fig:TSmap2}
\end{center}
\end{figure*}

\begin{figure}
\begin{center}
\includegraphics[width=0.49\textwidth]{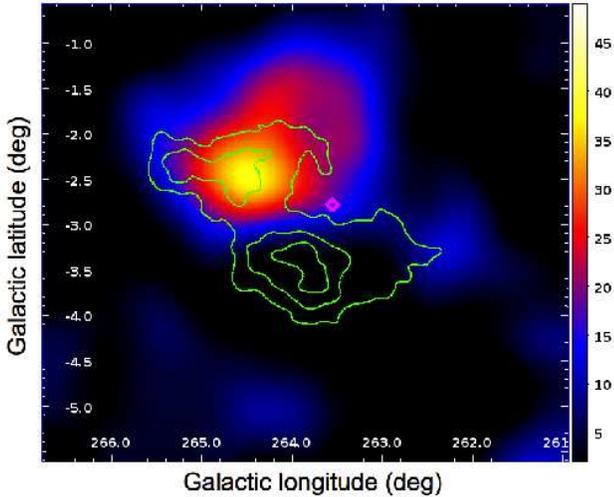}
\caption{\Fermi-LAT TS map  of the \VelaX\ PWN in the 0.3 -- 1.0 energy range (using $front$ events only, Galactic coordinates). The Southern wing of the radio emission has been included in the background model. Significant \gam-ray emission coincident with the Northern wing of the radio emission is detected by the \Fermi-LAT. The contours of the WMAP 61~GHz radio (in green) are overlaid for comparison. The position of the Vela pulsar is marked with a magenta diamond. }
\label{fig:TSmap3}
\end{center}
\end{figure}

We also examined the correlation of the \gam-ray emission from \VelaX\ with multi-wavelength observations of this PWN using $\mathtt{pointlike}$. We compared the TS obtained with the best-fit model, i.e. the elliptical Gaussian distribution, with the TS obtained using the templates derived from WMAP (61 GHz radio image, shown by green contours in Figure~\ref{fig:TSmap1}) and H.E.S.S. observations. For each analysis, a power law spectrum was assumed. The resulting TS values, which are equivalent to $2 \Delta(log(L))$, are summarized in the second half of Table~\ref{table:centroid_pointlike}. When comparing the results obtained by modeling the Fermi-LAT emission with multi-wavelength templates, using the H.E.S.S. template significantly decreases the value of the likelihood with respect to the WMAP template, as noted in the first publication reporting the \Fermi-LAT detection of the \VelaX\ PWN \citep{VelaX1}. However, we still observe a good correlation between the radio and the \Fermi-LAT observations. We also divided the radio template into two halves, as indicated in Figure~\ref{fig:TSmap1}, to look for an energy-dependent morphological behavior. The split radio model provides an improvement at $\sim$~3.6$\sigma$ and $\sim$~6.5$\sigma$ levels with respect to the single radio template and the Southern radio wing model respectively, and is also confirmed by the spectral analysis (see Section~\ref{subsection:spectrum}). 

The multi-wavelength templates and the analytical models cannot be compared directly since the models are not nested. In the following analysis we decided to use the best geometrical morphology implemented in $\mathtt{pointlike}$, namely the elliptical Gaussian distribution.

\begin{deluxetable*}{lcccccc}
\tablecaption{Centroid and extension fits to the LAT data for \VelaX\ using $\mathtt{pointlike}$ for $front$ events, assuming an elliptical Gaussian distribution. \label{table:energyvsmorpho} }
\tablewidth{0pt}
\tablehead{ 
\colhead{Energy range} & \colhead{R.A. ($^\circ$)} & \colhead{Dec. ($^\circ$) } & $\sigma_1$ ($^\circ$) & $\sigma_2$ ($^\circ$)& P.A. ($^\circ$) &   \colhead{$2 (log(L_{1})- log(L_{0}))$}
}
\startdata
0.3 -- 100 GeV			& 128.40 $\pm$ 0.06 & -45.40 $\pm$ 0.05 	& 0.46 $\pm$ 0.05 	& 1.04 $\pm$ 0.09	& 40.3 $\pm$ 4.0  	& 481	\\
0.3 -- 1.0 GeV			& 128.75 $\pm$ 0.11 & -45.33 $\pm$ 0.12 	& 0.88 $\pm$ 0.15	& 1.02 $\pm$ 0.16	& 44.4 $\pm$ 13.4 	& 216	 \\
1.0 -- 100.0 GeV 			& 128.26 $\pm$ 0.06 & -45.39 $\pm$ 0.06	& 0.43 $\pm$ 0.06	& 1.07 $\pm$ 0.11 	& 44.3 $\pm$  1.9	& 291	 \\
\enddata
\end{deluxetable*}

Figure~\ref{fig:TSmap2} presents the \Fermi-LAT TS maps of \gam-ray emission around the \VelaX\ PWN in two energy bands (0.3~--~1~GeV, 1~--~100.0~GeV) using $front$ events only. Radio and TeV contours have been overlaid for comparison. We attempted to characterize the energy-dependent shape of the PWN by estimating the source extension in these energy intervals. The centroids and extensions in the different energy ranges are summarized in Table~\ref{table:energyvsmorpho}. From Figure~\ref{fig:TSmap2} we note that the emission in the ``Northern wing'' (defined with respect to the Galactic coordinates) of the radio emission is bright in the lower energy band and becomes faint above 1 GeV, which might be an indication of a softer spectrum than the ``Southern wing''. 

It is worth noting that we report here for the first time the detection of \gam-ray emission from the Northern wing of the \VelaX\ PWN. This detection is clearly visible in the TS map presented in Figure~\ref{fig:TSmap3}, in which the Southern radio wing was included in the background model. Table~\ref{table:centroid_pointlike} shows that the log-likelihood of the fit is significantly improved by using the split radio templates instead of the Southern radio wing only. The discovery was enabled by the low energy threshold (0.3 GeV) now considered in this analysis. In addition, the extension and position of the Southern wing are in full agreement with the results of the morphological fit performed above 0.8 GeV and reported in the first \Fermi-LAT paper on \VelaX\ \citep{VelaX1}.

\subsection{Spectrum}\label{subsection:spectrum}
The following spectral analyses are performed with $\mathtt{gtlike}$ using $front$ and $back$ events between 0.2 and 100 GeV. We used the best morphological model from Table~\ref{table:centroid_pointlike}, i.e. the elliptical Gaussian distribution, to represent the \gam-ray emission observed by the LAT, as discussed in Section~\ref{subsection:morpho}. 

\begin{figure}
\begin{center}
\includegraphics[width=0.4\textwidth]{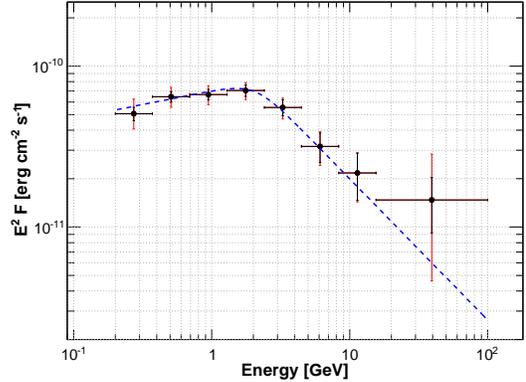}
\caption{Gamma-ray spectra of the \VelaX\ PWN, using the elliptical Gaussian spatial model. The blue dashed line shows the fit of a smoothly broken power law to the overall spectrum derived from all of the data with energy above 0.2 GeV. The data points (crosses) indicate the fluxes measured in each of the ten energy bins indicated by the extent of their horizontal lines. The statistical errors are shown in black, while the red lines take into account both the statistical and systematic errors as discussed in Section~\ref{subsection:spectrum}.  A 99.73\% C.L. upper limit is computed when the statistical significance is lower than 3$\sigma$.}
\label{fig:Spectra}
\end{center}
\end{figure}

Assuming this spatial shape, the \gam-ray source observed by the LAT is detected with a TS of 940 above 0.2 GeV. The spectrum of \VelaX\ above 0.2 GeV is presented in Figure~\ref{fig:Spectra}.  It is well described by a smoothly broken power law : 
\begin{equation}
\label{EQ:dNdE}
F(E)\,=\,\frac{dN}{dE}\,=\,N_0\,\left(\frac{E}{E_0}\right)^{-\Gamma_1}\left(1+\left(\frac{E}{E_b}\right)^{\frac{\Gamma_1-\Gamma_2}{\beta}}\right)^{-\beta}
\end{equation}
where $\Gamma_1$ = 1.83 $\pm$ 0.07 $\pm$ 0.27, $\Gamma_2$ = 2.88 $\pm$ 0.21 $\pm$ 0.06 are the spectral indices below and above the break energy $E_b$ = 2.1 $\pm$ 0.5 $\pm$ 0.5 GeV. The parameter $\beta$ is fixed to the value 0.2 as in standard \Fermi-LAT analyses \citep[e.g.][]{CrabFlare2012}. The first error is statistical, while the second represents our estimate of systematic effects as discussed below. The integrated flux renormalized to the total phase above 0.2 GeV is (1.83 $\pm$ 0.08 $\pm$ 0.25) $\times$ 10$^{-7}$ cm$^{-2}$~s$^{-1}$. This spectral model is favored over the simple power law and an exponential cut-off power law at 6.6$\sigma$ and 2.7$\sigma$ levels respectively. This is in agreement with results obtained independently using  $\mathtt{pointlike}$.  Similar results are obtained with the radio template, as can be seen in Table~\ref{table:spectra}. 

The \Fermi-LAT spectral points shown in Figure~\ref{fig:Spectra} were obtained by dividing the 0.2~--~100 GeV range into 10 logarithmically-spaced energy bins and performing a maximum likelihood spectral analysis to estimate the photon flux in each interval, assuming a power-law shape with fixed photon index $\Gamma$ = 2 for the source. The normalizations of the diffuse Galactic and isotropic emission were left free in each energy bin.  A 99.73\% C.L. upper limit is computed when the statistical significance is lower than 3$\sigma$. Bins at the highest energies corresponding to upper limits were combined.

\begin{figure*}
\begin{center}
\includegraphics[width=0.45\textwidth]{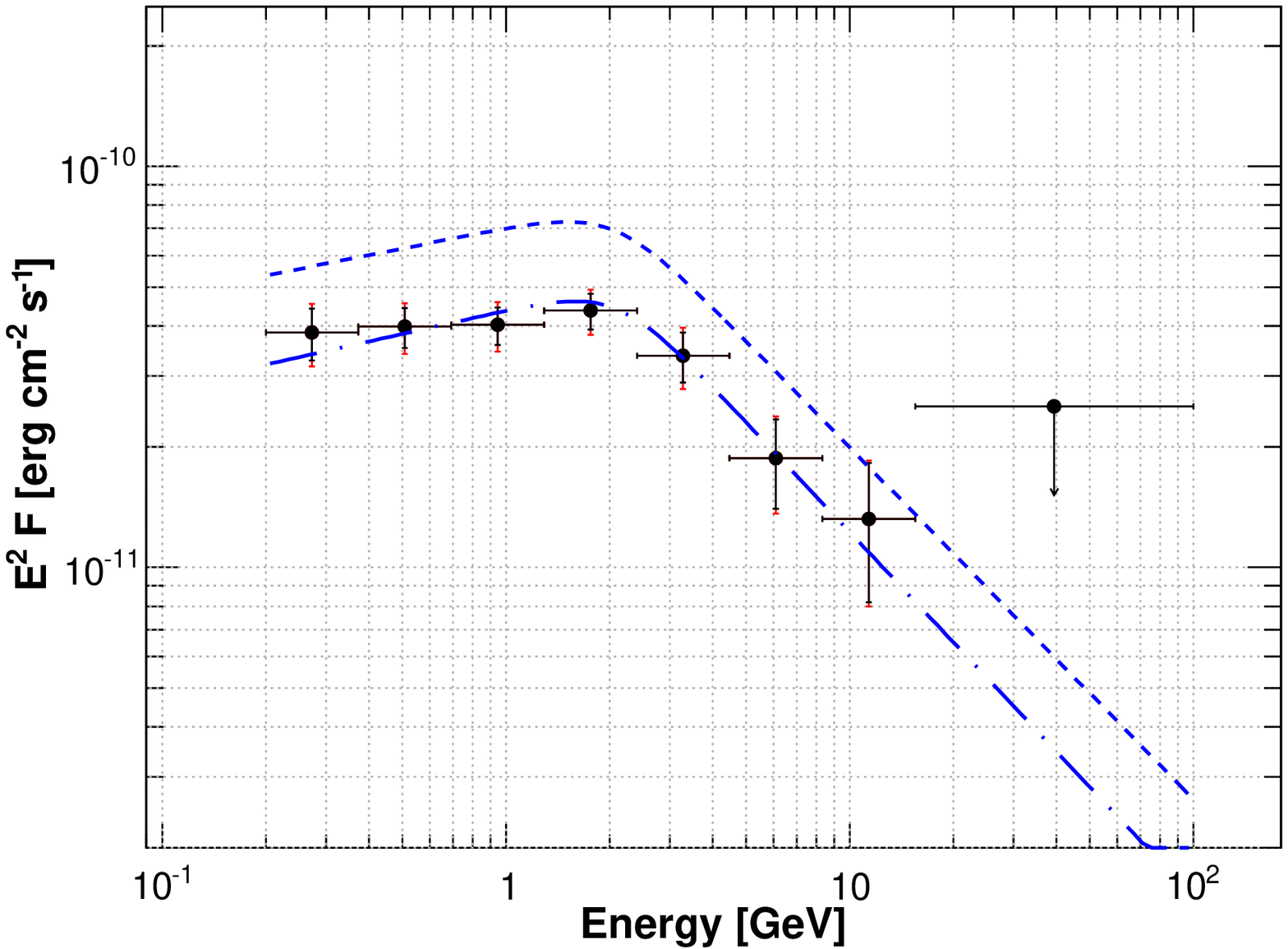}\includegraphics[width=0.45\textwidth]{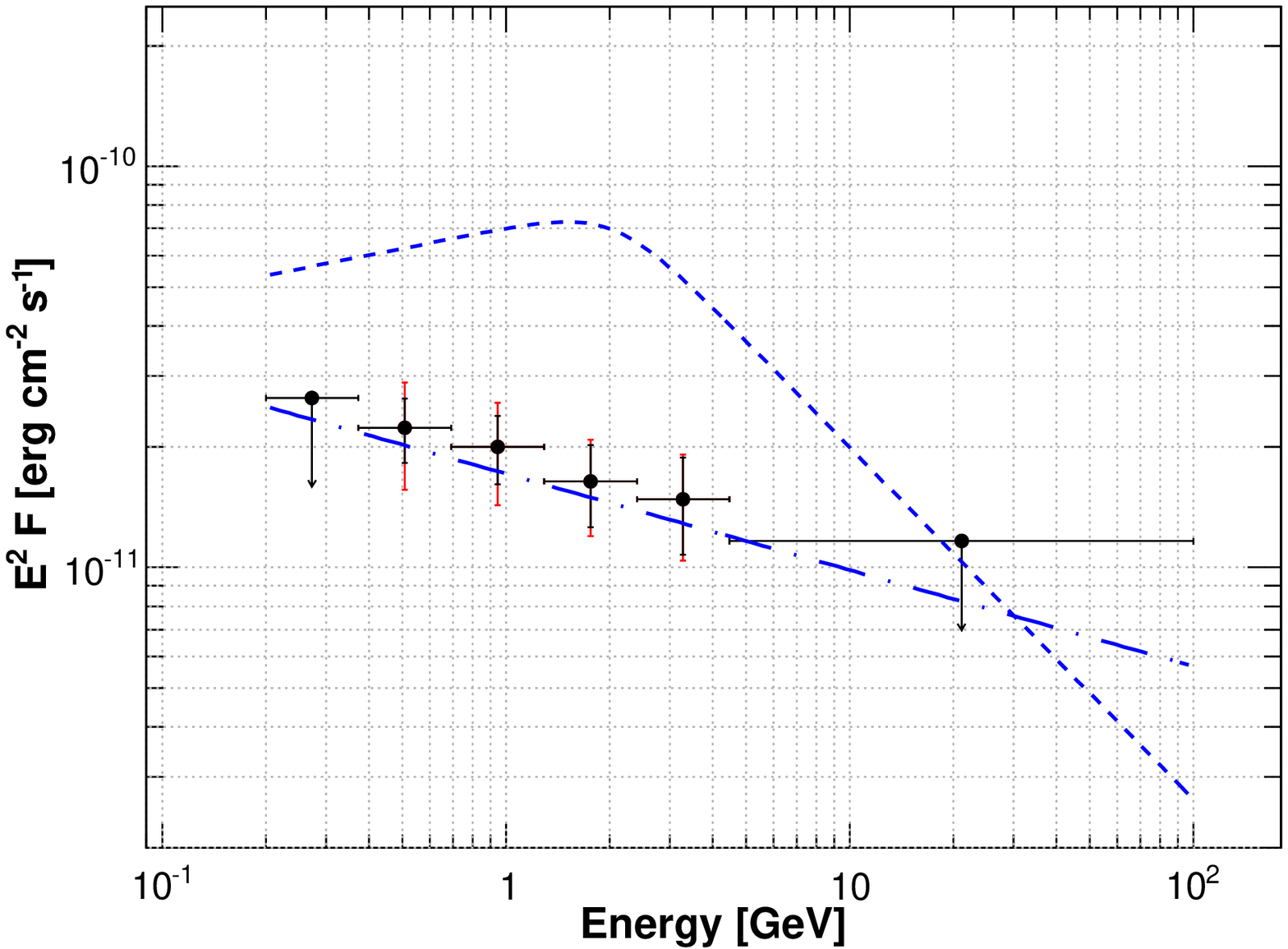}
\caption{Gamma-ray spectra of the two components of the \VelaX\ PWN, as defined in Section~\ref{subsection:morpho}. The blue dot-dashed line shows the best fit of the overall spectrum derived from all of the data with energy above 0.2 GeV. Left and Right images correspond to the Southern and Northern wings respectively. The spectrum obtained with the single radio template is indicated by a blue dashed line (as shown in Figure~\ref{fig:Spectra}). Plot conventions are similar to Figure~\ref{fig:Spectra}.}
\label{fig:SpectraHalf}
\end{center}
\end{figure*}

\begin{deluxetable*}{lccccc}
\tablecaption{Best spectral fit values obtained with $\mathtt{gtlike}$ using different templates for \VelaX\ above 0.2~GeV. The first and second errors denote statistical and systematic errors, respectively.\label{table:spectra} }
\tablewidth{0pt}
\tablehead{
 \colhead{Spatial Model} & \colhead{Flux [$10^{-7}$ ph cm$^{-2}$ s$^{-1}$]*} & \colhead{Photon index $\Gamma_1$} & \colhead{Photon index $\Gamma_2$} & Break energy $E_b$ & \colhead{TS}}
\startdata
Elliptical Gaussian &  1.83 $\pm$ 0.08 $\pm$ 0.25 & 1.83 $\pm$ 0.07 $\pm$ 0.27 & 2.87 $\pm$ 0.21  $\pm$ 0.06 & 2.1 $\pm$ 0.5 $\pm$ 0.5 & 940\\
\vspace{0.2cm} Radio &  1.73 $\pm$ 0.08 $\pm$ 0.22 & 1.88 $\pm$ 0.08  $\pm$ 0.27 & 2.89 $\pm$ 0.23 $\pm$ 0.05 & 2.0 $\pm$ 0.5 $\pm$ 0.5  & 897 \\
Split Radio templates : &  & &\\
\hspace{0.2cm} Northern wing**    & 0.64 $\pm$ 0.08 $\pm$ 0.14 & 2.25 $\pm$ 0.07  $\pm$ 0.20 & -- &-- & 116 \\
\vspace{0.2cm} \hspace{0.2cm} Southern wing  &  1.12 $\pm$ 0.08 $\pm$ 0.09 & 1.81 $\pm$ 0.10  $\pm$ 0.24 & 2.90 $\pm$ 0.25 $\pm$ 0.07 & 2.1 $\pm$ 0.5 $\pm$ 0.6 & 606 \\
\hline
$Disk11m$*** & 1.58 $\pm$ 0.07 & 2.24 $\pm$ 0.04 & --  &-- & 770 \\
$Disk11m$*** & 1.50 $\pm$ 0.07 & 1.96 $\pm$ 0.07 & 3.01 $\pm$ 0.30 &  2.0 $\pm$ 0.3 & 804\\
\enddata
\tablenotetext{*}{Fluxes are estimated above 0.2 GeV and renormalized to the total phase interval. The spectral parameter $\beta$ was fixed to the value 0.2.}
\tablenotetext{**}{The spectral parameters of the Northern wing were obtained after 2 iterations, as explained in the text.}
\tablenotetext{***}{The ``$Disk11m$'' model here refers to the best morphological fit obtained with 11 months of \Fermi-LAT data \citep{VelaX1}.}
\end{deluxetable*}

Four different systematic uncertainties can affect the LAT flux estimation : uncertainties in the Galactic diffuse background, in the morphology of the LAT source, in the effective area and in the energy dispersion. The fourth one is relatively small  ($\le$~10\%) and has been neglected in this study.  The main systematic at low energy is due to the uncertainty in the Galactic diffuse emission since \VelaX\ is located only $\sim$~3\deg\ from the Galactic plane. Different versions of the Galactic diffuse emission, generated by GALPROP~\citep{Strong2004},  were used to estimate this error. The observed \gam-ray intensity of nearby source-free regions on the Galactic plane is compared with the intensity expected from the Galactic diffuse models. We adopted the strategy described in \cite{w49} to estimate the expected intensity of the Galactic diffuse emission for different models. The difference, namely the local departure from the best-fit diffuse model, is found to be $\le 6$\%.  By changing the normalization of the Galactic diffuse model artificially by $\pm 6$\%, we estimate the systematic error on the integrated flux and on the spectral index. The second systematic is related to the morphology of the LAT source. The fact that we do not know the true \gam-ray morphology introduces another source of error that becomes significant when the size of the source is larger than the PSF. Different spatial shapes have been used to estimate this systematic error: a disk, a Gaussian distribution and the radio template. 
The third uncertainty, common to every source analyzed with the LAT data, is due to the uncertainties in the effective area.  This systematic is estimated by using modified instrument response functions (IRFs) whose effective area bracket that of our nominal IRF. These ``biased'' IRFs are defined by envelopes above and below the nominal dependence of the effective area with energy by linearly connecting differences of (10\%, 5\%, 10\%) at log(E) of (2, 2.75, 4) respectively.  We combine these various errors in quadrature to obtain our best estimate of the total systematic error at each energy and propagate them through to the fit model parameters.

Table~\ref{table:spectra} summarizes the obtained fluxes and spectral indices for each of the spatial templates described in Table~\ref{table:centroid_pointlike}.  

Using the elliptical Gaussian distribution and the smoothly broken power law and assuming a distance of $D = 290$~pc, the $\gamma$-ray luminosity of \VelaX\ above 0.2~GeV is $L_{\gamma}$ $\approx$ 2.4 $\times$10$^{33}$ ($D / 290$ pc)$^2$ erg s$^{-1}$, yielding a \gam-ray efficiency of $\eta = L_{\gamma}/\dot{E}$ =  0.03~\% of the spin-down power of the Vela pulsar. \\

We attempted to characterize the energy-dependent morphology of the \VelaX\ PWN by fitting the spectra associated with each of the split radio templates with independent spectral models. The results are summarized in Table~\ref{table:spectra}. The Northern wing is well modeled with a simple power law of index 2.25 $\pm$ 0.07 $\pm$ 0.20 while the Southern wing is better described by a smoothly broken power law. The corresponding spectral parameters are the following : $\Gamma_1$ =  1.81 $\pm$ 0.10  $\pm$ 0.24 and $\Gamma_2$ = 2.90 $\pm$ 0.25 $\pm$ 0.07 , with a break energy of $E_b$ = 2.1 $\pm$ 0.5 $\pm$ 0.6 GeV. Because of a potential interdependence of the wing spectral fits, the Northern parameters were obtained after 2 iterations. In a first step, both wings were fitted simultaneously. The Northern wing being much fainter than the Southern one, the spectral parameters of the Northern wing were re-adjusted in the second step, using fixed parameters for the Southern wing. Both iterations yield consistent results within statistical errors. However, the fit and spectral points obtained in the second step for the Northern wing are much more in agreement with each other. 
It is worth noting that the flux of the Northern wing is approximately half of the one in the Southern wing. However the Northern wing is located closer to the Galactic plane (i.e. in a region with a larger contribution from the Galactic diffuse background), which renders the emission from this wing less than half as significant as than the Southern wing. The improvement for the split radio model with respect to the single radio template is at $\sim$~4$\sigma$ level, which is consistent with the fact that the integral fluxes of the two radio wings are significantly different (see Table~\ref{table:spectra}). Figure~\ref{fig:SpectraHalf} (left and right) presents the spectra of the two regions modeled with the two split templates. Interestingly, this analysis shows that below $\sim$~2~GeV, the Northern wing has a softer spectrum by an index of $\sim$~0.5 with respect to the Southern wing, confirming the first indications given by the TS maps (see Figure~\ref{fig:TSmap2}). However, it should be noted that the steep spectrum of the Northern wing is mainly constrained by the upper limits at high energy. In this context, the likelihood of the fit is improved by only 2.5$\sigma$ when using a free power-law model instead of a broken power-law with fixed energy break and spectral indices (frozen at the values obtained for the Southern wing). More statistics are therefore needed to confirm spectral differences between the Northern and Southern regions. 

The careful reader may note that the best spectral fit of 4 years of \Fermi-LAT data (this paper) is obtained with a smoothly broken power law, while the fit of the 11 months of data yielded a simple power law of index $\Gamma$~$\sim$~2.4~$\pm$~0.1 and a weaker flux as presented in \cite{VelaX1}. These differences arise from the three main improvements (described below) made in this new analysis. 

First, a larger data set now enables us to spatially model the \VelaX\ \gam-ray emission with an elliptical Gaussian distribution, i.e. a more elaborate morphology than the``$Disk11m$'' model considered in the previous publication. The smaller extension of the $Disk11m$ with respect to the elliptical Gaussian distribution above 0.2 GeV therefore yields a fainter flux integrated over 0.2 GeV. For comparison, the spectral parameters obtained by fitting the 4-year dataset with a power law and a smoothly broken power law using the $Disk11m$ spatial model are included in Table~\ref{table:spectra}. 

Secondly, the increased statistics now allow a significant detection of a spectral break at $\sim$~2.0 GeV in the \gam-ray domain, which was not possible with only 11 months of data. Using $Disk11m$, the smoothly broken power law is preferred to the simple power law at 5.3$\sigma$ level. 

Finally, using the $Disk11m$ model, the harder spectrum obtained with the 4-year dataset (spectral index of $\Gamma$ = 2.24 $\pm$ 0.04 for a power law, see the first row labelled ``$Disk11m$'' in Table~\ref{table:spectra}) with respect to the 11-month dataset (which yielded a spectral index of $\Gamma$~=~2.4 $\pm$ 0.1) presented in \cite{VelaX1} likely arises from the slight contamination of the 11-month dataset by the Vela pulsar at low energies (below 1 GeV), which was estimated to be $\sim$~6\% of the \VelaX\ flux.  Defining the off-pulse window as 20\% of the pulsar phase (instead of 30\% in \cite{VelaX1}) ensures that we do not suffer contamination from the Vela pulsar in the new analysis.

\subsection{Multi-wavelength data}\label{subsection:radio}

Spectral measurements at different frequencies may help to better understand the origin of the emission observed from a source via the modeling of its spectral energy distribution. Considering the strong connection between the radio domain and the GeV energy range emphasized in \cite{VelaX1}, we examined in particular the data obtained in radio in this complex region.

\begin{figure}
\begin{center}
\includegraphics[width=0.45\textwidth]{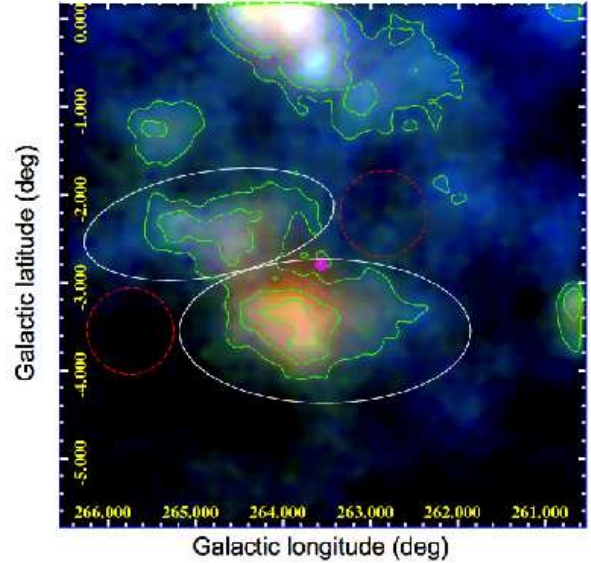}
\caption{Composite radio sky map obtained from WMAP data in Galactic coordinates (red : 41 GHz, green :  61 GHz, blue : 94 GHz) smoothed with a 3.6$^{\prime}$ Gaussian kernel. The magenta diamond shows the pulsar position. The extraction regions delimited with the white solid ellipses and the dashed red circles are used for the flux measurements for the source and the background respectively.The 61 GHz WMAP radio contours (0.80, 0.95 and 1.1 mK) are overlaid  as green solid lines. }
\label{fig:WMAPmap}
\end{center}
\end{figure}

Seven-year all-sky data of the \textit{Wilkinson Microwave Anisotropy Probe} (WMAP) were used to extract the spectrum of the \VelaX\ PWN at high radio frequencies. Five bands were analyzed, with effective central frequencies of 23, 33, 41, 61 and 94 GHz \citep{Jarosik2011}. Figure~\ref{fig:WMAPmap} represents the composite radio sky map in the \VelaX\ field of view. This image also shows evidence for a contaminating high frequency component superimposed on the PWN, especially on the Western side of the Northern wing, for Galactic coordinates (l, b) $\sim$ (265.8$^{\circ}$, -2.5$^{\circ}$). 

\begin{figure}
\begin{center}
\includegraphics[width=0.45\textwidth]{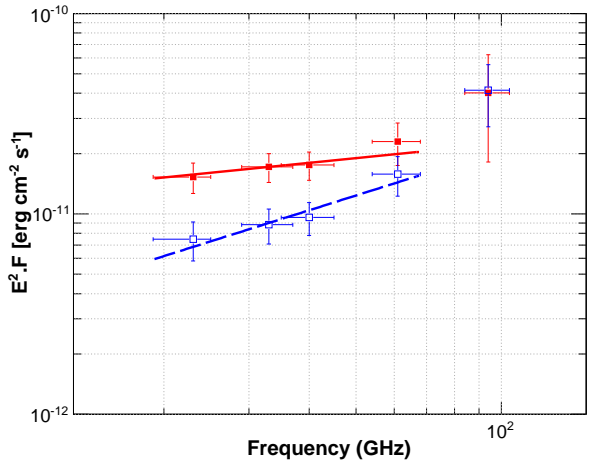}
\caption{Radio spectrum of the \VelaX\ PWN in the Southern (red, full squares) and Northern (blue, open squares) regions. The solid red and long dashed blue lines represent the best fit obtained in the 19~--~70~GHz frequency range for the Southern and Northern wings respectively.}
\label{fig:WMAPspectra}
\end{center}
\end{figure}

We extracted the radio spectral measurements using the regions delimited with the white ellipses and red circles for the source and background estimation respectively. The radio spectral points obtained within the Southern and Northern regions in the five frequencies covered by WMAP are summarized in Table~\ref{tab:SpectraRadio} and represented in Figure~\ref{fig:WMAPspectra}. When excluding the highest frequency spectral point (which may be contaminated as shown in Figure~\ref{fig:WMAPmap}), the fluxes in the two regions are well modeled with power laws $S_{i} \propto \nu^{-\alpha_i}$ (with $i =$ $S, N$ for the South and North respectively) of indices $\alpha_S$ = 0.76 $\pm$ 0.15 and $\alpha_N$ = 0.41 $\pm$ 0.08. Given the contamination noted above, this radio slope difference between the two wings may not be significant. Additional data, e.g. from Planck, and especially measurements above 40 GHz, are required to confirm this result.

\begin{deluxetable}{cccc}
\tablecaption{WMAP Spectral points corresponding to the Southern and Northern regions.\label{tab:SpectraRadio} }
\tablewidth{0pt}
\tablehead{
 \colhead{Band} & \colhead{Frequency} & \multicolumn{2}{c}{Flux}\\
 \colhead{} & \colhead{(GHz)} & \multicolumn{2}{c}{($\times$10$^{11}$~erg~cm$^{-2}$~s$^{-1}$)}\\
 \colhead{} & \colhead{} & \colhead{\textit{Southern region}} & \colhead{\textit{Northern region}} }
\startdata
$K$ & 23 & 1.53 $\pm$ 0.26 &	0.75$\pm$ 0.16 \\
$Ka$ & 33 & 1.72 $\pm$ 0.28 &      0.88$\pm$ 0.18 \\
$Q$ & 40 &1.76 $\pm$ 0.28 &      0.96$\pm$ 0.18 \\
$V$ & 60&  2.29 $\pm$ 0.55 &	1.58$\pm$ 0.35 \\
$W$ & 93 & 4.02 $\pm$ 2.20 & 	4.14$\pm$ 1.42 \\
\enddata
\end{deluxetable}

To better constrain the physical parameters related to the PWN and its environment, we will use the above described spectral measurements in the following section. A low-frequency (0.4 GHz) spectral point is extracted from \cite{Haslam1982} and will be used as an upper limit on the flux in each wing of the radio emission.

\section{Discussion}\label{section:discussion}
\subsection{Constraining the magnetic field in the halo extended region}
Determining the mechanism responsible for \gam-ray emission is crucial in order to measure the underlying relativistic particle population accelerated in a PWN. This new analysis confirms the excellent correlation between the GeV and the radio morphologies, showing that the \gam-ray emission extends well outside the narrow cocoon detected in X-rays and VHE. 

In a first step, we attempted to reproduce the multi-wavelength spectral energy distributions of the Southern and Northern wings of \VelaX\ without taking into account the emissions in the cocoon (detected in X-rays and TeV) which could be produced by a separate electron population, as explained in \cite{VelaX1}. The objective is to constrain the total energy injected in the form of electrons in the halo as well as the mean value of the magnetic field in this extended region. For this purpose, we used a one-zone model similar to the one described in \cite{1825}. The hadronic scenario, according to which the VHE emission is produced by proton-proton interactions and neutral pion decays, was proposed by \cite{Horns2006} but seems to be disfavored because of the large particle density required with respect to the density derived from X-ray observations \citep{LaMassa2008}. Therefore, it will be disregarded in the following. The spectral differences between the Northern and the Southern wing being marginal with the current statistics of the WMAP and \Fermi-LAT data, we did not try to reproduce this effect in the following scenario. In addition, due to the contamination visible at high frequencies in the WMAP data, we did not try to fit the spectral point at 94 GHz. 

We assume the same leptonic spectrum injected in each region. The WMAP spectral index of $\sim0.4$ in the Southern region (which is the less contaminated) requires a particle index of $\gamma \sim0.4 \times 2 + 1 \approx 1.8$, kept fixed in our model (see Table~\ref{table:fixpar}). In both regions, the leptonic spectrum injected shows an energy cut-off at the highest energy which is fitted and constrained by the observational data. 

Electrons suffer energy losses due to ionization, bremsstrahlung, synchrotron processes and inverse Compton (IC) scattering. Escape outside the halo is also taken into account assuming Bohm diffusion. The modification of the electron spectral distribution due to such losses is determined following \cite{Aharonian1997}. The electron population is evolved over the estimated lifetime of the pulsar (11 kyr). We fix here the pulsar braking index to the canonical value of $n$ = 3.0. The magnetic field and spin-down power of the pulsar are assumed to remain constant throughout the age of the system and do not depend on the size of the PWN. Neither the interaction of the reverse shock nor the diffusion of the leptons within the PWN are modeled, since there are not enough observables to sufficiently constrain their corresponding parameters. In this context, this phenomenological model is used to reproduce the multi-wavelength data assuming that the Vela pulsar is the only source of energy.

The IC photon field includes the cosmic microwave background (CMB), far infrared from the dust (IR; temperature of 25 K, density of 0.44 eV~cm$^{-3}$) and starlight (Optical; temperature of 7500 K, density of 0.44 eV~cm$^{-3}$), reasonable for the locale of \VelaX\ \citep{DeJager2008}. We assume a distance of $D = 290$~pc and a size for each region (i.e. the Northern and Southern wings) of $10$~pc. 

\begin{figure*}
\begin{center}
\includegraphics[width=0.7\textwidth]{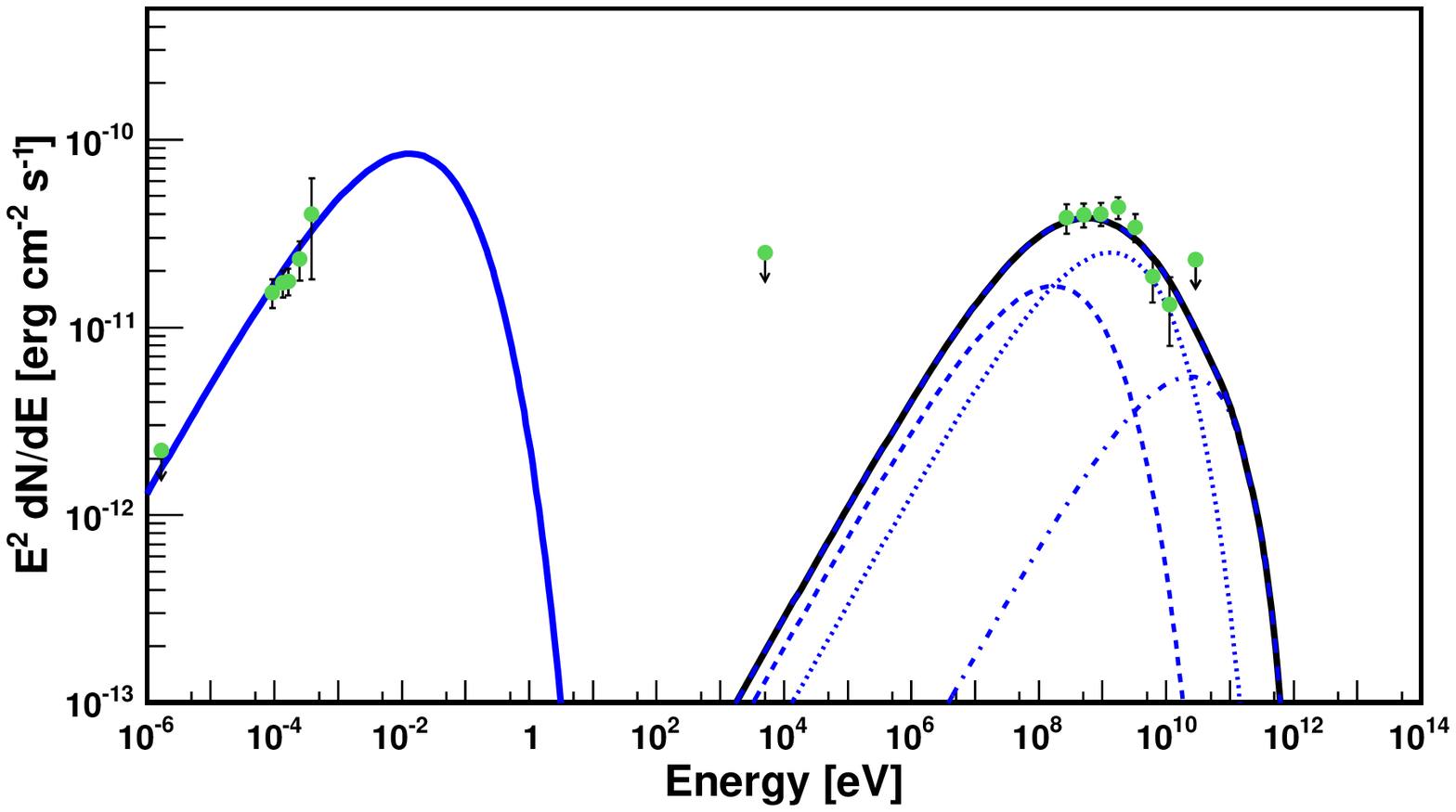}
\includegraphics[width=0.7\textwidth]{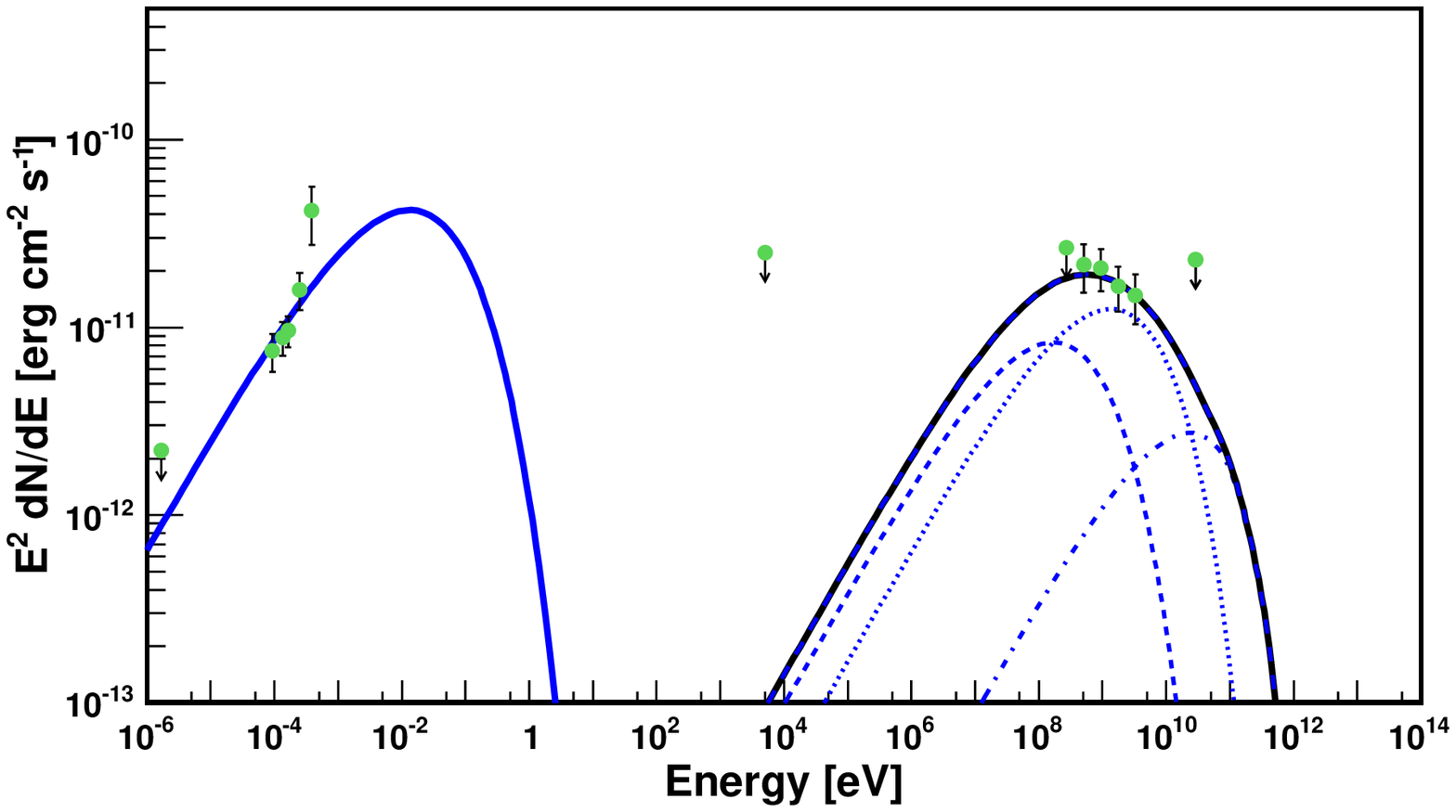}
\caption{Spectral energy distributions of the Southern (top) and Northern (bottom) wings from radio to \gam-rays. WMAP and \Fermi-LAT spectral points (this paper) are represented with green points. The ROSAT upper limit \citep{VelaX1} is also shown. The low frequency radio upper limit is derived from \cite{Haslam1982}. The dashed, dotted and dot-dashed lines represent the inverse Compton components from scattering on the CMB, dust emission and starlight respectively. The sum of the three \gam-ray components is shown as a solid curve.}
\label{fig:modeling}
\end{center}
\end{figure*}

All in all, our simple one-zone model has four free parameters adjusted to reproduce the photon spectral energy distribution as seen in radio \citep[][this work]{Haslam1982} and \gam-rays (this work) in both regions: the magnetic field $B$ in the halo, the exponential cut-off energy (which are both assumed to be the same in the Northern and Southern regions) and the fraction $\eta$ of the pulsar spin-down power injected to particles in each region. The best fit, obtained by minimizing the $\chi^2$ statistic between model and data points, is shown in Figure~\ref{fig:modeling} (top and bottom). As can be seen in this figure, a simple power-law injection model can reasonably well reproduce the multi-wavelength data of the two wings. The best-fit parameters are summarized in Table~\ref{tab:modeling}.
Model fitting is achieved by minimizing the $\chi^2$ between model and data using the simplex method described in \cite{Nelder1965}. This algorithm is included within the ROOT framework provided by the CERN\footnote{ROOT : http://root.cern.ch/}. For each ensemble of N variable parameters we evolve the system over the pulsar lifetime and calculate the $\chi^2$ between model curves and flux data points. The simplex routine subsequently varies the parameters of interest to minimize the fit statistic. We estimate parameter errors by using MINOS, which is designed to calculate the correct errors in all cases, especially when there are non-linearities. The theory behind the method is described in \cite{Eadie1971}.
It is worth noting that $\sim$26\% and $\sim$13\% of the total energy injected by the pulsar (which represents 100\% of the spin-down power injected in particles accounting for all losses during the lifetime of the system) is required to power the radio to \gam-ray emission from the Southern and Northern wings respectively. In addition, the total energy injected into leptons and the magnetic field derived are both in very good agreement with previous estimates \citep{DeJager2008, Aharonian2006, VelaX1}. The synchrotron/IC peak ratio of the cocoon implies a magnetic field of 4~$\mu$G with very small uncertainty, which can be compared to our value of $\sim 5 \mu$G in the halo. Since no data are available to trace the synchrotron peak and better constrain the magnetic field in the halo, we cannot exclude similar values in the halo and in the cocoon. 

\subsection{Rapid diffusion of electrons ?}
The radio emission, arising from synchrotron radiation, traces the magnetic field distribution. On the other hand, in a leptonic scenario, the HE emission is produced via IC scattering and directly traces the underlying relativistic electron distribution. The new \Fermi-LAT results together with the correlation between the radio and \gam-ray data now provide direct evidence that low energy electrons are present in the extended halo. The recent H.E.S.S. detection of TeV emission coincident with the extended radio halo \citep{Abramowski2012} is additional evidence that electrons are present in this large structure. In addition to that, the above simple modeling provides further evidence that the magnetic fields in the halo and cocoon regions are not strongly different. 
In that case, the real puzzle is to understand the origin of the electrons present in the extended radio halo since significant emission is now detected by \Fermi-LAT and H.E.S.S. up to $\sim 1^{\circ}$ from the Vela pulsar, i.e $\sim 10$ pc from the powering pulsar.

A potential scenario was first proposed by \cite{vanetten2011} to explain the large size of the PWN HESS~J1825$-$137: rapid diffusion of high energy particles with $\tau_{\rm esc} \sim 90 (R/10 \, \rm{pc})^2 (E_e/100 \, \rm{TeV})^{-1} $ year (where $R$  and $E_e$ are the radius of the PWN and the energy of the injected electrons respectively) which is 1000 times faster than standard Bohm diffusion. This is in contradiction with the common assumption of toroidal magnetic fields with strong magnetic confinement. The authors argue that turbulence and mixing caused by the passage of the reverse shock might provide the necessary disruption to the magnetic field structure to allow particles to diffuse far more rapidly. More recently, this fast diffusion was invoked for \VelaX\ by \cite{Hinton2011} to interpret the steep \Fermi-LAT spectrum and the absence of $> 100$ GeV photons in the extended radio halo. Their best fit to the data yields a factor of 2000 enhancement over Bohm diffusion. In such a scenario, the \gam-ray flux observed by \Fermi-LAT at the outer realm of the \VelaX\ extended nebula would be produced by high energy electrons that were injected when the pulsar was much younger. However, their model does not produce any TeV emission at large distance from the pulsar since high energy electrons escape too fast. A way to solve this issue would be to inject two populations of electrons \citep[as suggested earlier by ][]{VelaX1} and decrease the diffusion time so that 10 TeV photons can still be visible up to $1^{\circ}$ (10 pc) from the powering pulsar in a magnetic field $B$ of 5~$\mu$G. Since the standard Bohm diffusion time scale is $\tau_{\rm diff} \sim 34 (R/1 \, \rm{pc})^2 (E_e/10 \, \rm{TeV})^{-1}(B/10 \, \mu\rm{G}) $ kyr \citep{Zhang2008}, a diffusion only $\sim$20 times faster than Bohm diffusion would be needed. It should be noted that this scenario should lead to spectral differences between the inner and the outer regions of the PWN due to radiative cooling of the electrons during their propagation, which is in contradiction with the recent H.E.S.S. results \citep{Abramowski2012}. Energy-dependent diffusive escape and stochastic re-acceleration in the radio halo could explain the absence of spectral variations in the TeV regime but such complex modeling is out of the scope of our paper.

Another possibility would be that these electrons do not come from the Vela pulsar and are directly accelerated in this extended structure through stochastic acceleration due to turbulent magnetic fields in the outer PWN flow or in the surrounding SNR plasma, since there is no evidence of shocks in this region. Such 2$^{nd}$-order Fermi acceleration accounts well for the radio emission from supernova remnants~\citep{Scott1984}. It provides an excess of accelerated electrons that will radiate through synchrotron and IC radiation as seen in radio by WMAP and \gam-rays by \Fermi-LAT. The maximum energy to which electrons can be accelerated by such mechanism depends highly on the level of turbulence, which is unknown. To reproduce the \Fermi-LAT spectrum, a reasonable maximum energy of $\sim$140 GeV is required. Obviously, in this case, the radio/\Fermi\ halo would not be linked with the X-ray/TeV cocoon. The TeV extended halo, if related to the radio structure, would need a second component of accelerated electrons to reproduce the peaked \gam-ray spectrum. Unfortunately, the comparison of the radio, GeV and TeV emissions is limited by the differences in resolution and sensitivity of the instruments involved. In particular, the current multi-wavelength observations do not allow conclusions to be drawn about whether the TeV and radio emissions arise from the exact same location.

\section{Conclusion}\label{section:conclu} 
Using four years of \Fermi-LAT observations and a lower energy threshold for the morphological analysis than previously, we report for the first time the detection of \gam-ray emission from the Northern wing of the \VelaX\ PWN. The best-fit geometrical morphological model in the 0.3~--~100~GeV energy range is obtained for an elliptical Gaussian distribution. We also report the detection of a significant energy break at $E_b$ = 2.1 $\pm$ 0.5 GeV in the \Fermi-LAT spectrum as well as a marginal spectral difference between the Northern and the Southern wings. WMAP data have also been used to characterize the synchrotron emission in the two wings of the radio halo. However, the WMAP image shows evidence for a contaminating high frequency component superimposed on the PWN, especially on the Northern wing, and the radio slope difference observed between the two wings may not be significant.

Further observations to characterize the radio and \gam-ray spectra are required and will help understand the origin of the excess of low energy electrons detected in these two wavelengths. High frequency radio observations are clearly needed to spatially resolve the radio emission, especially above 60 GHz. Increased sensitivity with the continued observations by \Fermi-LAT will also enable any spectral differences between the Northern and Southern regions to be firmly established. In addition, observations with H.E.S.S.-II will provide a better overlap with \Fermi-LAT and therefore a direct link to verify if the TeV signal is related to the extended structure detected by \Fermi. In addition, deeper observations with high sensitivity instruments such as \textit{XMM}-Newton will help to better constrain the spectra and their potential spatial variations in the cocoon and halo in the X-ray domain. Despite a large sample of multi-wavelength data, \VelaX\ remains an excellent case to study and the future observations in radio and \gam-rays will obviously provide new clues to understand the acceleration mechanisms taking place in this complex object, as well as new surprises.\\
 
\begin{deluxetable*}{lc}{ht}
\tablecaption{Values of the parameters assumed for the modeling. \label{table:fixpar} }
\tablewidth{0pt}
\tablehead{ 
\colhead{Parameters} & \colhead{Value}
}
\startdata
Spectral index $\gamma$				&	1.8		\\
Age (kyr) 						& 11 \\
Spin-down power $\dot{E}$	(erg~s$^{-1}$)	& 6.9 $\times$ 10$^{36}$	\\
Pulsar braking index $n$ & 3.0 \\
\enddata
\end{deluxetable*}

\begin{deluxetable*}{lcc}{ht}
\tablecaption{Best-fit parameters for the Southern and Northern wings. \label{tab:modeling} }
\tablewidth{0pt}
\tablehead{ 
\colhead{Parameters} & \colhead{Southern wing} & \colhead{Northern wing}
}
\startdata
Magnetic field B  ($\mu$G) 			&	\multicolumn{2}{c}{4.9 $\pm$ 0.8}	\\
Cut-off energy (GeV)						&	\multicolumn{2}{c}{145 $\pm$ 30}	\\
Fraction of spin-down power $\eta$ (\%) 	&	26 $\pm$ 5 	& 	13 $\pm$ 2	\\
Total energy injected to leptons ( $\times$~10$^{47}$ erg)		&	6.2 $\pm$ 1.2	& 	3.1 $\pm$ 0.5	\\
\enddata
\end{deluxetable*}

%
\noindent{\it Acknowledgements}\\
\small
The \textit{Fermi} LAT Collaboration acknowledges generous ongoing support from a number of agencies and institutes that have supported both the development and the operation of the LAT as well as scientific data analysis. These include the National Aeronautics and Space Administration and the Department of Energy in the United States, the Commissariat \`a l'Energie Atomique and the Centre National de la Recherche Scientifique / Institut National de Physique Nucl\'eaire et de Physique des Particules in France, the Agenzia Spaziale Italiana and the Istituto Nazionale di Fisica Nucleare in Italy, the Ministry of Education, Culture, Sports, Science and Technology (MEXT), High Energy Accelerator Research Organization (KEK) and Japan Aerospace Exploration Agency (JAXA) in Japan, and the K.~A.~Wallenberg Foundation, the Swedish Research Council and the Swedish National Space Board in Sweden.

Additional support for science analysis during the operations phase is gratefully acknowledged from the Istituto Nazionale di Astrofisica in Italy and the Centre National d'\'Etudes Spatiales in France.

MHG acknowledges support from the Alexander von Humboldt Foundation. MHG wishes to thank Felix A. Aharonian and Patrick O. Slane for helpful comments and discussions. 


\begin{thebibliography}{}
\bibitem[{Abdo et al., 2009a}]{Vela1}  Abdo, A. et al., 2009a, ApJ, 696, 1084
\bibitem[{Abdo et al., 2009b}]{Abdo2009b}  Abdo, A.  et al., 2009b, Phys. Rev. D, 80, 122004
\bibitem[{Abdo et al., 2010a}]{Vela2}  Abdo, A.  et al., 2010a, ApJ, 713, 154
\bibitem[{Abdo et al., 2010b}]{VelaX1}  Abdo, A. et al., 2010b, ApJ, 713, 146
\bibitem[Abdo et al., 2010c]{MSH1552} Abdo, A., et al., 2010c, ApJ, 714, 
\bibitem[Abdo et al., 2010c]{w49} Abdo, A. et al., 2010c, ApJ, 722, 1303
\bibitem[Abramowski et al., 2012]{Abramowski2012} Abramowski, A. et. al., 2012, A\&A, 548, A38
\bibitem[Ackermann et al., 2012]{Ackermann2012} Ackermann, M. et al., 2012, ApJS, 203, 4
\bibitem[Aharonian et al., 1997]{Aharonian1997} Aharonian, F.~A., Atoyan, A.~M. \& Kifune, T., 1997, MNRAS, 291, 162
\bibitem[Aharonian et al., 2006]{Aharonian2006} Aharonian, F.~A. et al., 2006, \aap, 448, L43 
\bibitem[Alvarez et al., 2001]{Alvarez2001} Alvarez, H., et al. 2001, A\&A, 372, 636
\bibitem[Atwood et al., 2009]{Atwood 2009} Atwood, W. et al., 2009, ApJ, 697, 2, 1071
\bibitem[Beringer et al., 2012]{Beringer2012} Beringer, J. et al. (Particle Data Group), 2012, Phys. Rev. D86, 010001
\bibitem[Buehler et al., 2012]{CrabFlare2012} Buehler, R. et al. 2012, ApJ, 749, 26
\bibitem[Caraveo et al., 2001]{Caraveo2001} Caraveo, P. A., De Luca A., Mignani R. P. \& Bignami, G. F. 2001, ApJ, 561, 930
\bibitem[de Jager et al., 1996]{DeJager1996} de Jager, O.~C., Harding, A.~K., Strickman, M.~S., ApJ, 460, 729
\bibitem[de Jager et al., 2008]{DeJager2008} de Jager, O.~C., Slane, P.~O., LaMassa S.~M., 2008, ApJL, 689, L125
\bibitem[Dodson et al., 2003]{Dodson2003} Dodson, R., et al. 2003, ApJ, 596, 1137
\bibitem[Eadie et al., 1971]{Eadie1971} Eadie, W.~T.,  Drijard, D., James, F., Roos, M. and Sadoulet, B. , 1971, Statistical Methods in Experimental Physics, North-Holland, 1971, pp. 204-205
\bibitem[Fang \& Zang, 2010]{Fang2010} Fang, J. \& Zhang, L., 2010, A\&A, 515, 20
\bibitem[Gaensler \& Slane, 2006]{Gaensler2006} Gaensler, B. M. \& Slane, P. O., 2006, ARA\&A, 44, 17
\bibitem[Grondin et al., 2011]{1825} Grondin, M.-H. et al, 2011, ApJ, 738, 42
\bibitem[Haslam et al., 1982]{Haslam1982} Haslam, C. G. T., Salter, C. J., Stoffel, H. \& Wilson, W. E. 1982, A\&AS, 47,1
\bibitem[Hewitt et al., 2012]{PuppisA} Hewitt, J., Grondin, M.-H., Lemoine-Goumard, M., Reposeur, T. et al, 2012, ApJ, 759, 89
\bibitem[Hinton et al., 2011]{Hinton2011} Hinton, J., Funk, S., Parsons, R.~D. \& Ohm, S., 2011, ApJL, 743, L7
\bibitem[Helfand et al., 2001]{Helfand2001} Helfand, D. J., Gotthelf, E. V., \& Halpern, J. P. 2001, ApJ, 556, 380
\bibitem[Hobbs et al., 2006]{Hobbs2006} Hobbs, G. B., Edwards, R. T., \& Manchester, R. N., 2006, MNRAS, 369, 655
\bibitem[Horns et al., 2006]{Horns2006} Horns, D., Aharonian, F., Santangelo, A., Hoffmann, A.~I.~D., \& Masterson, C.\ 2006, \aap, 451, L51 
\bibitem[Jarosik et al., 2011]{Jarosik2011} Jarosik, N., et al. 2011, ApJS, 192, 14
\bibitem[Kanbach et al., 1994]{Kanbach1994} Kanbach, G. et al., 1994, \aap, 289, 855
\bibitem[Kerr, 2011]{Kerr2011} Kerr, M. 2011, PhD Thesis, arXiv:1101.6072v
\bibitem[LaMassa et al., 2008]{LaMassa2008} LaMassa, S. M., Slane, P. O., \& De Jager, O. C. 2008, ApJ, 689, L121
\bibitem[Lande et al., 2012]{Lande2012} Lande, J. et al., 2012, ApJ, 756, 5
\bibitem[Large et al., 1968]{Large1968} Large, M.~I., Vaughan, A.~E., and Mills, B.~Y., 1968, \nat, 220, 340
\bibitem[Markwardt  \& \"Ogelman(1995)]{Markwardt1995} Markwardt, C.~B., \& \"Ogelman, H.\ 1995, \nat, 375, 40 
\bibitem[Mattox et al., 1996]{Mattox1996}   Mattox, J.~R. et al., 1996, ApJ, 461, 396
\bibitem[Nelder \& Mead, 1965]{Nelder1965} Nelder, J.~A. \& Mead, R., 1965, Comput. J., 7, 308
\bibitem[Nolan et al., 2012]{SecondCat} Nolan, P.~L. et al. 2012, ApJS, 199, 31
\bibitem[\"Ogelman et al., 1993]{Ogelman1993} \"Ogelman, H., Finley, J.~P. and Zimmerman, H.~U., 1993, Nature, 361, 136
\bibitem[Pellizzoni et al., 2010]{Pellizzoni2010} Pellizzoni, A. et al., 2010, Science, 327, 663
\bibitem[Ray et al., 2011]{Ray2011} Ray, P.~S., et al., 2011, ApJS, 194, 17
\bibitem[Reynolds et al., 2012]{Reynolds2012} Reynolds, S.~P., Gaensler, B.~M. \& Bocchino, F., 2012, Space Science Reviews, 166, 231
\bibitem[Rishbeth, 1958]{Rishbeth1958} Rishbeth, H., 1958, Australian Journal of Physics, 11, 550
\bibitem[Scott \& Chevalier, 1984]{Scott1984} Scott, J. S., Chevalier, R. A., 1975, Astrophys. J. Lett. 197, L5
\bibitem[Slane et al., 2010]{Slane2010} Slane, P.O. et al., 2010, ApJ, 720, 266
\bibitem[Slane et al., 2012]{Slane2012} Slane, P.O., et al., 2012, ApJ, 749, 131
\bibitem[Spitkovsky, 2008]{Spitkovsky2008} Spitkovsky, A., 2008, ApJL, 682, L5
\bibitem[Strong et al., 2004]{Strong2004} Strong, A. W., Moskalenko, I. V., \& Reimer. O. 2004, ApJ, 613, 962
\bibitem[Tanaka et al., 2011]{VelaJr} Tanaka, T. et al., 2011, ApJL, 740, L51
\bibitem[Thompson, 1975]{Thompson1975} Thompson, D.~J., Fichtel, C.~E., Kniffen, D.~A., \"Ogelman, H.~B., 1975, ApJ, 200, L79
\bibitem[Van Etten \& Romani, 2011]{vanetten2011} Van Etten, A., \& Romani, R. W., 2011, ApJ, 742, 62
\bibitem[Wallace et al., 1977]{Wallace1977} Wallace, P.~T. et al., 1977, \nat, 266, 692
\bibitem[Weiler \& Panagia, 1980]{Weiler1980} Weiler, K. W., Pangia, N., 1980, A\&A, 90, 269 
\bibitem[Zhang et al., 2008]{Zhang2008} Zhang, L, Chen, S.~B., \& Fang, J., 2008, ApJ, 676, 1210
\end{thebibliography}
\end{document}